\documentclass{Interspeech}
\usepackage{multirow}

\interspeechcameraready

\title{Training Articulatory Inversion Models for Interspeaker Consistency\thanks{This research is funded by an EPSRC DTP and the Vice-Chancellor's Award. It is supported by Cambridge University Press \& Assessment, a department of The Chancellor, Masters, and Scholars of the University of Cambridge.}}

\author[affiliation={1}]{Charles}{McGhee}
\author[affiliation={1}]{Mark J.F.}{Gales}
\author[affiliation={1}]{Kate M.}{Knill}

\affiliation{ALTA Institute/MIL, Department of Engineering}{University of Cambridge}{UK}

\email{cgm43@cam.ac.uk,  mjfg@eng.cam.ac.uk, kmk1001@cam.ac.uk}


\usepackage{comment}

\begin{document}

\maketitle

\begin{abstract}
    Acoustic-to-Articulatory Inversion (AAI) attempts to model the inverse mapping from speech to articulation. Exact articulatory prediction from speech alone may be impossible, as speakers can choose different forms of articulation seemingly without reference to their vocal tract structure. However, once a speaker has selected an articulatory form, their productions vary minimally. Recent works in AAI have proposed adapting Self-Supervised Learning (SSL) models to single-speaker datasets, claiming that these single-speaker models provide a universal articulatory template. In this paper, we investigate whether SSL-adapted models trained on single and multi-speaker data produce articulatory targets which are consistent across speaker identities for English and Russian. We do this through the use of a novel evaluation method which extracts articulatory targets using minimal pair sets. We also present a training method which can improve interspeaker consistency using only speech data.
    
\end{abstract}

\section{Introduction}

Recovering an arbitrary speaker's exact articulation from speech is a difficult, if not impossible, task.
Speakers can produce consonants, such as /\textturnr/ in American English, in different ways which give almost the same acoustic output and their production may not be completely determined by their vocal tract structure \cite{westbury1998differences}.
This suggests that even if we could reliably estimate vocal tract structure from speech, there could still be multiple plausible hypotheses of the articulatory movements behind any particular acoustic signal.
An inversion model trained to reduce error on multi-speaker paired acoustic-articulatory data might overfit speaker attributes in an attempt to model this ambiguity.
However, individual speakers are fairly consistent with their articulation for consonant and vowel production \cite{westbury1998differences,johnson1993individual}.
Therefore, a better approach may be to train an inversion model to produce outputs which match a consistent articulatory profile, mapping productions of articulatorily ambiguous but acoustically similar consonants like /\textturnr/ to a single movement.
The resulting inversion model would represent a canonical speaker in a manner similar to the canonical (or seed) parameters for speaker adaptation in Automatic Speech Recognition (ASR) \cite{woodland2001speaker}.

The idea of using a single speaker's articulatory data to create such a model was explored in \cite{cho2024coding}. 
The authors claimed that this model produced a universal articulatory encoding, justifying this claim with low Word Error Rates (WER) on a multilingual articulatory synthesis task.
Whilst articulatory synthesis should form part of the evaluation of these models, the combination of source and articulatory features can make it difficult to determine which part of the input is responsible for the resulting speech sounds.
One property of a universal articulatory encoding would be that articulatory targets are easily separable for the vowels and consonants of a given language.
If the encoding matches the canonical profile described above, we'd also expect that these targets to vary minimally across different speakers.
In this paper, we detail a model and reference-free evaluation method aimed at separating consonant and vowel sounds into their respective articulatory targets using minimal pair sets.
We'll then compare AAI models trained on single and multi-speaker datasets using this method and introduce a speech-only fine-tuning method to improve target separation.

\section{Minimal Pair Evaluation}

To examine how articulatory targets change for specific consonants and vowels, we first need to control for the surrounding phonetic environment.
We do so by comparing a set of words which all vary in terms of a single phoneme (a minimal pair set). 
We begin with an analysis of real articulatory data, providing a method to identify articulatory targets for the contrastive sounds in a minimal pair set.
We then demonstrate a method to find minimal pair sets from pronunciation dictionaries, before creating an evaluation set using speech synthesis.

\subsection{XRMB Evaluation}

\begin{figure}[!htp]
    \centering
    \includegraphics[width=.7\columnwidth]{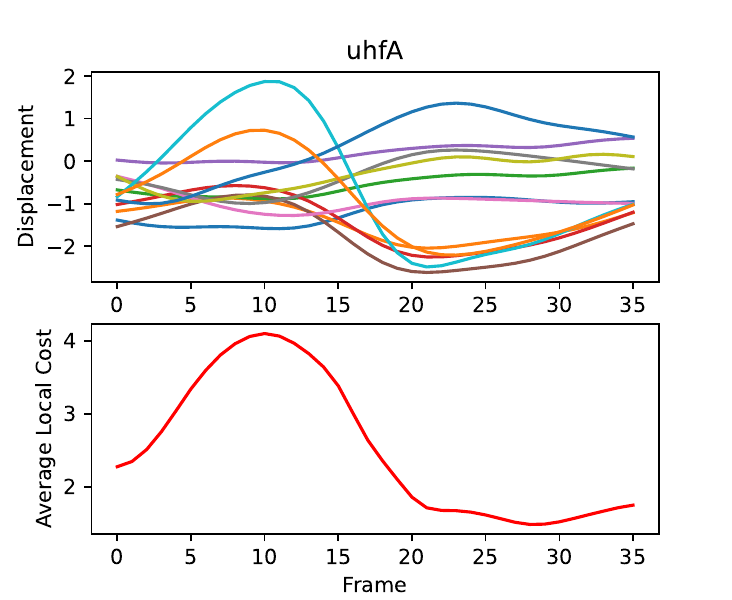}
    \caption{Articulatory trajectories (Top), average local DTW cost (Bottom) for the utterance `uhfA' from XRMB.}
    \label{dtw_uhfA}
\end{figure}

The X-Ray Microbeam dataset (XRMB) \cite{westbury1994x} contains a task (tp016) where speakers were asked to produce a number of vowel-consonant-vowel (V-C-V) samples changing only the inner consonant (e.g. ``uhfA", ``uhkA"). 
To find articulatory targets for these samples, we use Dynamic Time Warping (DTW) \cite{senin2008dynamic} to align the articulatory features in each V-C-V sample to the rest of that speaker's samples and find the point where the average local cost (Euclidean distance in this case) is maximised.
We do this for 18 speakers in the dataset who have samples that are not mistracked for the following phonemes in the C position: /f/, /k/, /t/, /\textesh/, /b/, /\textturnr/,  /w/,  /l/, /z/.
The articulatory points we align are the first three tongue positions (T1-3, where T1 is the tongue tip), lower lip, upper lip and central incisor x and y coordinates, z-normed by speaker.
We use these point-based features in the inversion experiments as well, in line with \cite{mcghee2024highly}.
The articulatory traces for ``uhfA" along with the average DTW local cost are plotted in Figure \ref{dtw_uhfA}. 

\begin{figure}[!htp]
    \centering
    \includegraphics[width=1\linewidth]{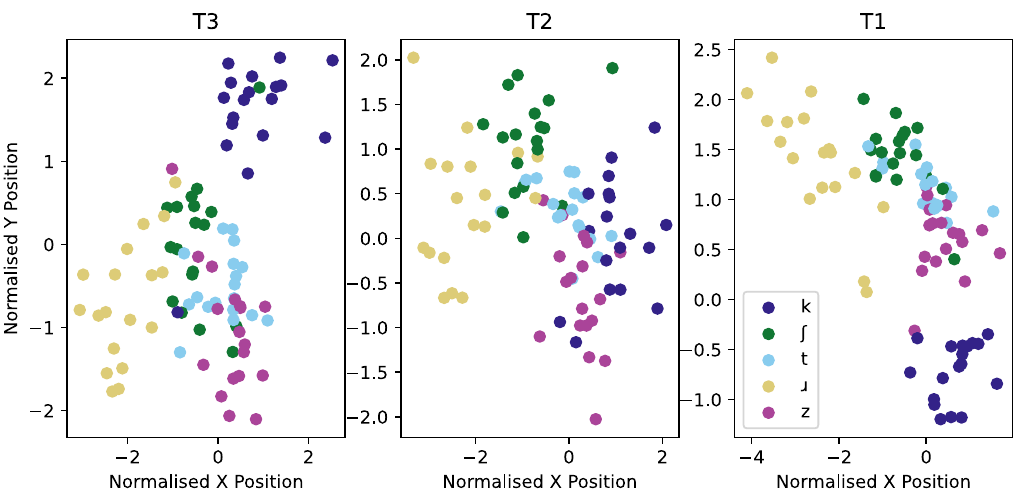}
    \caption{First three tongue positions (T1-3) for consonants from XRMB separated using DTW local cost maximum}
    \label{separated_consonants}
\end{figure}

Whilst the positions obtained through the local cost maximum may not be the absolute target, finding the nearest critical point (as in \cite{ananthakrishnan2011mapping}) requires that the articulatory trajectories be smooth.
Although we do impose some smoothing constraints on the analysis (a low-pass filter with a cutoff of 10Hz), we do not have a reliable method to detect when the trajectories have been oversmoothed.
We also find that using the maximum of the DTW local cost produces target-like positions for real and synthetic data.
For real data, this can be seen in Figure \ref{separated_consonants} where we plot the first three tongue positions for consonants which are largely discriminated by tongue position.

Whilst we align the points intraspeaker, if the model is consistent across speakers, the targets should vary minimally from one another and be easily separable.
We therefore perform classification across all the points using leave-one-out classification and a linear Support Vector Machine (SVM).
The accuracy of this method on the XRMB data detailed above was 92\%. The same alignment and classification procedure is used to evaluate the AAI models detailed in Section 3.

\subsection{Minimal Pair Sets}

We create the evaluation set using a pronunciation dictionary and speech synthesis.
We first define a graph where the vertices of the graph are the words in the pronunciation dictionary and each edge is a phoneme swap that defines a minimal pair according to that pronunciation dictionary.
Cliques (vertex subsets where every pair of vertices is adjacent) in this graph are minimal pair sets.
We look at two accents of English, UK (Standard Southern British English) and US (General American), and one accent of Russian (Moscow).
We choose Russian as it contains phonemically contrastive palatilization and a high central vowel that is not present in the accents of English we have chosen.
We use the \verb|english_us_mfa|, \verb|english_uk_mfa| and \verb|russian_mfa| pronunciation dictionaries from the Montreal Forced Aligner \cite{mcauliffe2017montreal} to define the graph and the Bron-Kerbosch algorithm \cite{bron1973algorithm} for finding cliques.
We then manually selected 10 minimal pair sets for each condition. 
5 of these sets were vowel minimal sets designed to cover the basic monophthongs of each accent/language.
The other 5 were consonant minimal sets, which contained a randomly selected version of each voiced/unvoiced consonant in the set.
To synthesise the minimal pairs, we use Google's \verb|Standard| speech synthesis model \footnote{\url{https://cloud.google.com/text-to-speech}}.
We choose this model as it synthesises single words with minimal prosodic variation whilst retaining distinct speaker attributes.
This set is used to evaluate the AAI models in Section 3.

\subsection{Voicing Score}

We also introduce a score which measures whether the features are acting as articulatory features or as acoustically discriminant features.
This voicing score takes one of the US consonant minimal pair sets (`bail', `fail', etc) from Section 2.2 and adds back in consonants which vary by voicing and nasality, but not place and manner (`pale' and `mail').
We then measure the maximum distance between the targets given by the model for these voicing/nasality distinct minimal pairs and compare this distance with every other consonant that differs by place and manner.
If there is a distance which is less than the maximum voicing/nasality distance, we mark this as an error (value of 0); otherwise, we mark the contrast as correct (value of 1). 
The final score is given by the number of correct samples divided by the total number of samples.
We tested this score using several random projections of WavLM features to the articulatory output dimension, finding the score for these random projections to range around 50\%.

\section{Articulatory Inversion}

The current state-of-the-art in AAI adapts WavLM either with a linear layer \cite{cho2024coding} or Low Rank Adaptation (LoRA) \cite{hu2021lora} fine-tuning \cite{mcghee2024highly}. We train our own variants of these models in this section and compare these with the model presented in \cite{cho2024coding}.

\subsection{Single and Multi-Speaker Models}

\begin{figure}[!htp]
    \centering
    \includegraphics[width=\columnwidth]{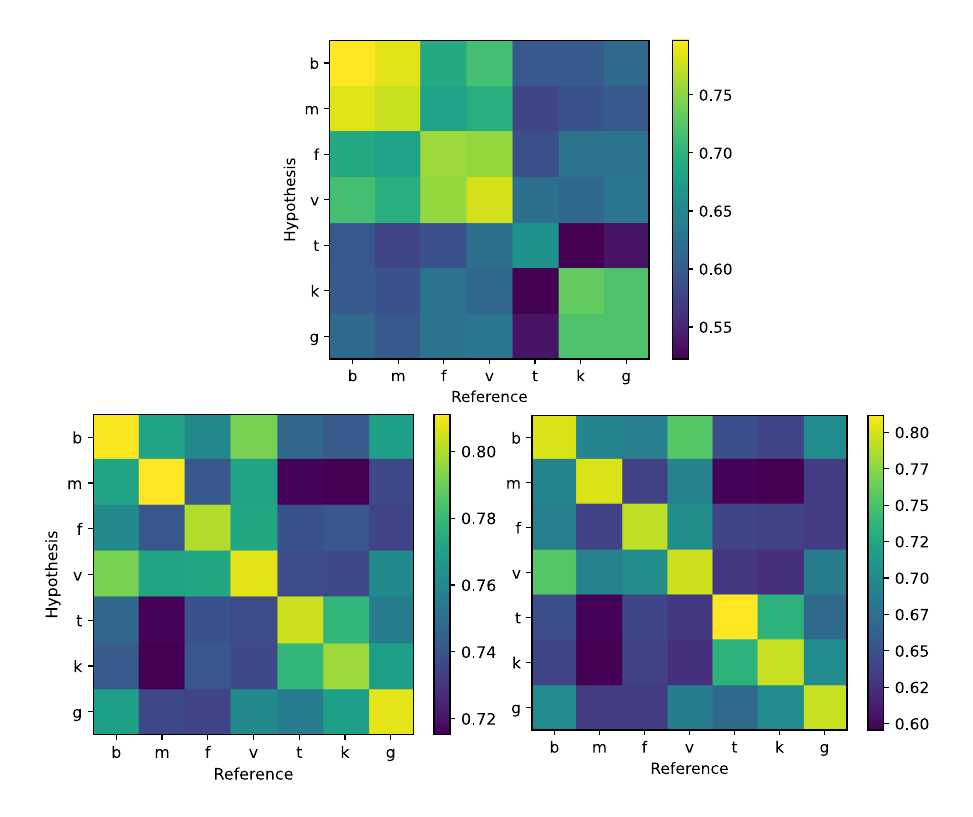}
    \caption{Average cosine similarities between DTW-aligned articulatory features (top), WavLM 9th layer (bottom left) and final layer features (bottom right) for phonemically-contrastive samples from XRMB.}
    \label{XRMB_truth}
\end{figure}

The single and multi-speaker datasets used in this study are the dataset from \cite{medina2022speech} (henceforth called Medina) and XRMB.
We use the whole dataset when training the single-speaker models and the train/validation split from \cite{mcghee2024highly} for training models on XRMB.
We adapt a WavLM Large model in two ways, firstly with a linear layer and secondly with LoRA.
Whilst there is some evidence that features from the intermediate layers of WavLM might perform better for inversion \cite{cho2023evidence}, we find that final-layer and intermediate layer features encode similar knowledge about the similarity between different consonants. 
We DTW aligned every combination of speaker (except the same speakers) and consonant from a subset of the XRMB data in Section 2.1 using a cosine distance cost function and present the average cosine similarities between DTW aligned features for each consonant combination in Figure \ref{XRMB_truth}.
The articulatory representations are more similar for contrasts which are similar in terms of place and manner (e.g. bilabials), whereas 9th and final layer WavLM features are acoustically discriminant, with voiceless plosives /t/ and /k/ being more similar than the voicing contrast /k/ and /g/.
We therefore opt to only use final layer features in the experiments as there is evidence that there is less speaker information in these layers \cite{chen2022does}.

Following \cite{mcghee2024highly} we use a One Cycle learning rate scheduler \cite{smith2019super} and a max learning rate of 1e-3 for 10 epochs.
The LoRA parameters are set to the same as \cite{mcghee2024highly}, with a rank of 32 and alpha of 16.
We use the average English minimal pair classification accuracy as an early stopping criterion (patience of 3) for all models, bar one LoRA model that is trained without early stopping.
All articulatory data is z-normed by speaker and 10Hz low-pass filtered using a 5th-order Butterworth filter.
We use a lower cutoff frequency in this study than in \cite{mcghee2024highly} to match results from \cite{cho2024coding}.
In Table \ref{big_results}, we give the minimal pair classification performance and voicing score, alongside the correlation and Mean Squared Error (MSE) for all models except the model from \cite{cho2024coding} as we did not have access to the dataset and therefore could not project the output points to the XRMB articulatory space.

\subsection{WavLM Guided Consistency Training}

To ensure the model output is the same for two speakers producing the same phonetic content, we use consistency regularisation \cite{sajjadi2016regularization}.
Consistency regularisation encourages a model to be resistant to irrelevant perturbations, such as noise, in the input that should not affect the model's output.
In this case, the perturbation is speaker identity so the sample $\textbf{x}$ and perturbed sample $\textbf{x}^*$ are two renditions of the same lexical content produced by two different speakers.
We use a consistency regularisation loss $L_{c}$ together with a self-training loss $L_{st}$ and weighting factor $\alpha$ to give a total loss, $L_{total}$:

\begin{equation}
    L_{total} = \alpha L_{st}  + (1 - \alpha) L_{c}
\end{equation}
\\
The self-training loss $L_{st}$ is given by: 

\begin{equation}
    L_{st} = \frac{1}{N}\sum_{i=1}^{N} \lvert\lvert \hat{\textbf{y}}_{i} - \textbf{y}_{i} \rvert\rvert^2
\end{equation}
\\
Where $\hat{\textbf{Y}}$ is the model prediction (length N) on the unperturbed utterance and $\textbf{Y}$ is a prediction from a copy of the same model that is frozen during training.
In designing the consistency loss $L_{c}$, we need to align the perturbed utterance $\textbf{x}^*$ to $\textbf{x}$ as they will likely be different lengths.
Each element of the loss should also be weighted by a factor which accounts for differences in pronunciation.
To achieve both, we perform a DTW alignment of WavLM final layer features for the input $\textbf{W}$ and perturbed sample $\textbf{W}^*$, giving a warping function $\phi$ which maps the time indices of $\textbf{W}$ to $\textbf{W}^*$ and a weighting factor $c$ which is the cosine similarity between two aligned WavLM features.
As the AAI models are WavLM-based, the output runs at the same frame-rate and we can use $\phi$ and $c$ to create the consistency loss $L_{c}$:

\begin{equation}
    L_{c} =\frac{1}{N}\sum_{i=1}^{N} c_{({i},\phi(i))}\lvert\lvert \hat{\textbf{y}}_{i} - \hat{\textbf{y}}^*_{\phi(i)} \rvert\rvert^2 
\end{equation}

Where $\hat{\textbf{Y}}^*$ is the model prediction on the perturbed utterance.
We create $\textbf{x}$ and $\textbf{x}^*$ using VCTK \cite{yamagishi2019cstr}, by randomly selecting a paired speaker for each utterance in the corpus that is produced by two or more speakers.
The consistency training is run for a single epoch using a One Cycle learning rate scheduler and a max learning rate of 1e-5.
We swept $\alpha$ from 0 to 1 in 0.25 increments, finding a value of 0.25 to perform the best.

\begin{table*}[htp]
\centering
\caption{Articulatory metrics, Mean Squared Error (MSE) and correlation on XRMB evaluation set, minimal pair scores and WER results. ES - Early Stopping, VS - Voicing Score, EN - English Average.}
\begin{tabular}{ccccccccccc}
\cline{3-11}
                             &                     & \multicolumn{2}{c}{\textbf{Articulatory}} & \multicolumn{5}{c}{\textbf{Minimal Pairs} $\uparrow$}                                   & \multicolumn{2}{c}{\textbf{Intelligibility (WER \%)} $\downarrow$} \\ \hline
\textbf{Training Set}        & \textbf{Adaptation} & \textbf{MSE} $\downarrow$     & \textbf{Correlation} $\uparrow$   & \textbf{VS}  & \textbf{US}   & \textbf{UK}   & \textbf{EN}   & \textbf{RU}   & \textbf{LibriTTS-R}        & \textbf{RUSLAN}       \\ \hline
\multirow{2}{*}{Medina}      & Linear              & \textbf{0.585}   & 0.668                  & 96.8         & 96.3          & 97.7          & 97.0          & 90.8          & 1.39                       & 4.59                  \\
                             & LoRA                & 0.610            & \textbf{0.681}         & 99.0         & 94.3          & 96.0          & 95.2          & 87.4          & 1.54                       & 5.62                  \\ \hline
mngu0 \cite{cho2024coding}                        & Linear (No ES)      & -                & -                      & 97.6         & 92.0          & 95.4          & 93.7          & 86.7          & 1.80                       & 4.99                  \\ \hline
\multirow{3}{*}{XRMB}        & Linear              & 0.331            & 0.807                  & 92.7         & \textbf{98.3} & 98.9 & \textbf{98.6} & \textbf{94.4} & \textbf{1.25}                       & \textbf{4.22}                  \\
                             & LoRA (No ES)        & \textbf{0.251}   & \textbf{0.865}         & 99.2         & 93.6          & 95.8          & 94.7          & 83.5          & 1.53                       & 5.69                  \\
                             & LoRA                & 0.297            & 0.837                  & \textbf{100} & 97.2          & 97.7          & 97.5          & 90.5          & 1.37                       & 5.19                  \\ \hline
\multirow{2}{*}{XRMB + VCTK} & Linear              & 0.345            & 0.800                  & 89.5         & 96.8          & \textbf{99.3}          & 98.1          & 93.6          & 1.34                       & 4.37                  \\
                             & LoRA                & 0.327            & 0.818                  & \textbf{100} & 97.9          & 98.3          & 98.1          & 86.5          & 1.31                       & 5.51                  \\ \hline
\end{tabular}
\label{big_results}
\end{table*}

\section{Articulatory Synthesis}

In addition to our minimal pair classification evaluation, we also test the intelligibility of speech produced by articulatory synthesis using the inversion output.

\subsection{Synthesis Architecture}

We follow \cite{mcghee2024highly} and use an 8-layer conformer which takes in articulatory features, log F0 and log energy.
We deviate from \cite{mcghee2024highly} and use PyWORLD to estimate F0, as we found it minimally improved intelligibility over CREPE \cite{kim2018crepe}.
We do not use any additional source embedding, such as the utterance-level embedding in \cite{cho2024coding}, since we aim to produce intelligible synthesis with as few confounding factors as possible.

\subsection{Data}

In \cite{mcghee2024highly}, it was found that this synthesis model can be quite sensitive to recording conditions due to its reliance on energy as a feature.
We therefore select datasets which have either been cleaned or contain clean TTS-quality data.
For evaluating English, we use LibriTTS-R \cite{koizumi23_interspeech}, which is a version of LibriTTS \cite{zen2019libritts} where the samples are identical but the sound quality has been improved through speech restoration.
We train each model on the LibriTTS-R train-clean-100 subset for 3 epochs using a One Cycle learning rate scheduler and a max learning rate of 1e-3.
To evaluate each model for intelligibility, we synthesise samples on the LibriTTS dev-clean set and use Whisper large-v3 \cite{radford2023robust} to generate transcripts for the original speech and synthetic speech.
We score the transcripts against each other, giving a cross WER which we report in Table \ref{big_results}.
To evaluate Russian, we opt for the RUSLAN \cite{gabdrakhmanov2019ruslan} dataset.
Although it is a single-speaker dataset, it provides a consistent acoustic environment that might not be present in a dataset like Multilingual Librispeech \cite{pratap2020mls}.
We use a 90/10 train/validation split and the same learning hyperparameters as the English data.
We also evaluate intelligibility in the same way as the English data, giving the cross WER in Table \ref{big_results}.

\section{Results}

\begin{figure}[!h]
    \centering
    \includegraphics[width=1\linewidth]{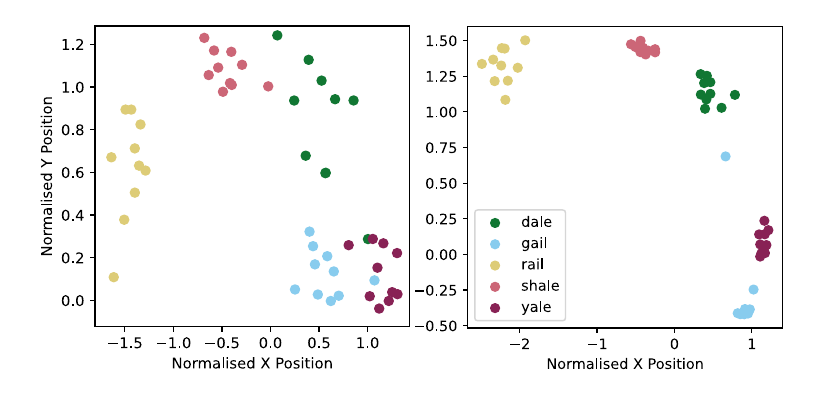}
    \caption{T1 (tongue tip) positions extracted for a subset of one of the UK minimal pair sets with features from the linear multi-speaker model (left) and consistency trained LoRA model (right).}
    \label{tongue_tip_wavlm_lora}
\end{figure}

From the results given in Table \ref{big_results}, it is clear that the multi-speaker models outperformed the single-speaker models in terms of minimal pair classification and intelligibility, with linear-adapted models performing marginally better than LoRA-adapted models.
All models achieved voicing scores that were much higher than the random WavLM projection benchmark of $\sim$50\%, showing that all models learnt a form of articulatory representation, although none of the linear models achieved 100\% voicing score.
A crucial result was that models which had higher correlation on the XRMB evaluation set did not perform better on the classification and intelligibility tasks.
Additionally, there was a strong negative correlation between the average minimal pair classification performance and WER, with correlations of -0.96 for the English and -0.89 for the Russian set.
We found in general that models which scored highly on the minimal pair classification task would output clear articulatory targets, as in Figure \ref{tongue_tip_wavlm_lora}.

\begin{figure}[!htp]
    \centering
    \includegraphics[width=\linewidth]{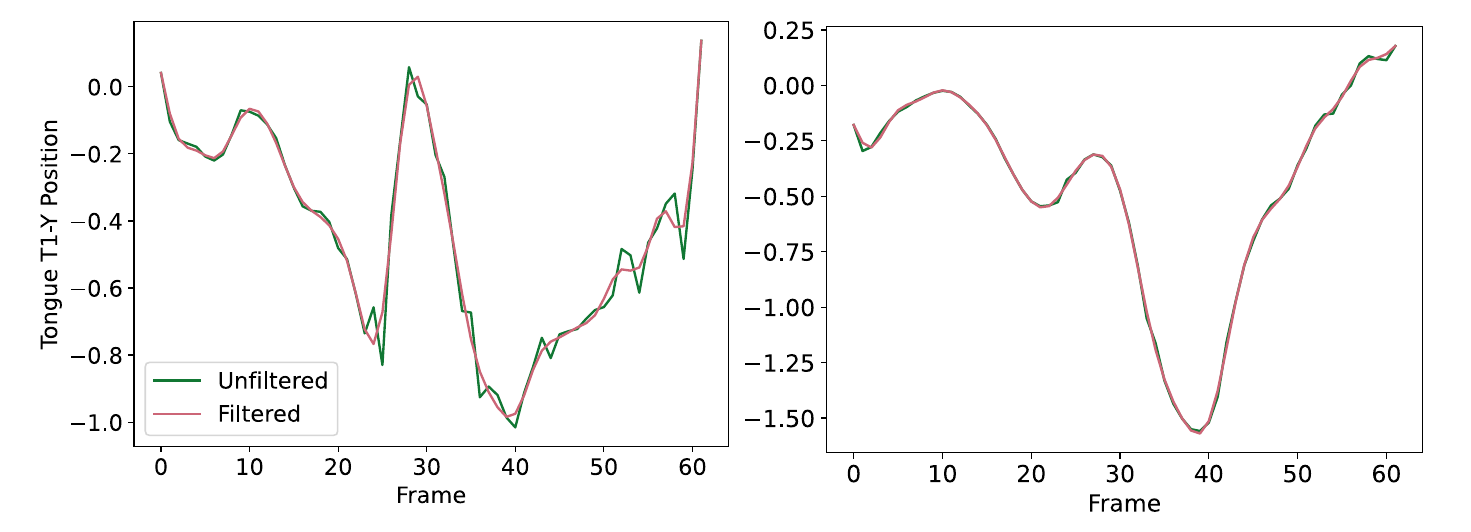}
    \caption{Filtered and unfiltered T1-Y position from linear multi-speaker model (left) and consistency trained LoRA model (right).}
    \label{smoothness}
\end{figure}

The target positions from linear models had noticeably higher variance than those from LoRA-adapted models.
We haven't examined the articulatory dynamics of these models in any detail, but the higher variance of the linear models is likely a result of the fact that the predicted trajectories are not smooth, as is shown in Figure \ref{smoothness}.
Although we use a low-pass filter in the minimal pair analysis, the lack of smoothness in the output trajectories may affect the final target positions.
As is also visible in Figure \ref{smoothness}, the LoRA-adapted models do produce smooth trajectories, but whether they are valid articulatory dynamics \cite{elie2023modeling} is a topic for future work.
Once a method is created for evaluating dynamics, we could also use critical point analysis \cite{ananthakrishnan2011mapping} to get the exact targets for classification.

The consistency loss generally improved classification accuracy for both models on British English, but only improved overall accuracy and intelligibility on both accents of English for the LoRA-adapted model.
It also reduced Russian classification accuracy and intelligibility for both models.
This indicates that the training method may have a bias towards the accent composition of the data the model is trained on, with VCTK containing a large amount of British English-accented speech.
In the future, we'll look to use speech synthesis to generate the perturbed utterance $\textbf{x}^*$ and apply this method to larger sets of multilingual data.
Classification accuracy and intelligibility results on Russian were significantly worse than for English across all the models. 
This indicates that these AAI models do not produce fully universal articulatory features.
However, we hope that this evaluation task can be used to iteratively improve performance on multilingual inversion.

\section{Conclusion}

In this paper, we have presented a model and reference-free method to evaluate synthetic articulation that showed strong correlations with intelligibility results on an articulatory synthesis task.
Using this evaluation method, we demonstrated that AAI models can easily overfit paired acoustic-articulatory data, multi-speaker models can outperform single-speaker models in terms of interspeaker consistency and SSL models adapted to English articulatory data do not fully generalise to languages like Russian.
We also introduced a new method for fine-tuning AAI models on only speech data, which improved target separation for LoRA-adapted models.
In the future, we'll expand the minimal pair evaluation and consistency loss to more languages to see whether we can build a universal AAI model.

\bibliographystyle{IEEEtran}
\bibliography{mybib}

\end{document}